# IMAGE SIMULATION FOR SPACE APPLICATIONS WITH THE SURRENDER SOFTWARE

**Jérémy Lebreton, Roland Brochard,**
**Matthieu Baudry, Grégory Jonniaux, Adrien Hadj Salah, Keyvan Kanani, Matthieu Le Goff, Aurore Masson, Nicolas Ollagnier, Paolo Panicucci, Amsha Proag, Cyril Robin**

*Airbus Defence & Space, 31 rue des Cosmonautes, 31402 Toulouse*
*surrender.software@airbus.com*
*https://www.airbus.com/space/space-exploration/SurRenderSoftware.html*

## ABSTRACT

Image Processing algorithms for vision-based navigation require reliable image simulation capacities. In this paper we explain why traditional rendering engines may present limitations that are potentially critical for space applications. We introduce Airbus SurRender software v7 and provide details on features that make it a very powerful space image simulator. We show how SurRender is at the heart of the development processes of our computer vision solutions and we provide a series of illustrations of rendered images for various use cases ranging from Moon and Solar System exploration, to in orbit rendezvous and planetary robotics.

## 1   INTRODUCTION

The simulation of space scenes presents specific challenges, which are typically not handled by general purpose image simulators. Vision-based navigation solutions require training and validation datasets that are as close as possible to real images. Our team and partners develop computer vision algorithms for space exploration (Mars, Jupiter, asteroids, the Moon), and for in-orbit operations (rendezvous, robotic arms, space debris removal). There is a new wave of missions targeting cislunar orbit or the Moon surface. Of course "real images" are rarely available before the mission. Ground-based test facilities such as robotic test benches embarking mock-ups or experiences with scaled mission analogues (mars terrain analogue, drones flights, etc.) are useful, yet they are limited. For example it is very difficult to capture the scale of space scenes in a room-sized facility (such as a small objects illuminated by an extended light source). Also limited numbers of images are available from previous missions or from lab experiments, when thousands are needed to represent the variety of possible configurations that an algorithm will encounter. Another decisive asset of computer simulation is that the ground-truth is perfectly known, whereas real-life experiments are prone to errors and biases, which are hard to estimate or lack accuracy.

Some of the effects visible in space images are not of particular importance for traditional image simulators. For example, for far-range rendezvous, very low SNR targets (SNR ~ 1) must be simulated with high radiometric fidelity. Space-qualified cameras often have unusual optical distortions and achromatism, which also vary with camera aging and temperature, and the geometrical performance relies on properly modelling them. The Point Spread Function (PSF) and associated effects (resolution, blooming) are fundamental parameters for image quality and they need to be simulated physically. Defocus is often encountered and shall be well simulated. In this paper, we provide a quick listing of available rendering engines and discuss their limitations for space computer vision applications. We show how Airbus SurRender software attempts to go beyond these limitations. It is at the heart of the development process for many image processing solutions mostly in VBN. We use it from early prototyping, to extensive performance test campaigns and hardware-in-the loop experiments. Various API (Python, Matlab, Simulink, C++, etc.) are available to interface SurRender with different simulations environments (GNC environment simulator, optical stimulators, etc.). In the context of a growing need for autonomy and artificial intelligence in space, our team is pursuing a constant effort in the development of the SurRender software and its diffusion. In this paper we provide a general presentation of the software and its performances that must be read in complement to the details already introduced before [1]. We show a series of recent use cases simulated for our projects ranging for Lunar and Solar System exploration to in-orbit rendezvous with artificial objects and planetary robotics.

## 2 SPECIFICITIES OF IMAGE SIMULATION FOR SPACE APPLICATION

### 2.1 Requirements

In sectors developing computer vision solutions such as big tech and automotive industries, algorithms are often trained on real data because it is relatively easy to assemble massive datasets. Doing so is not an option for space applications because acquiring representative images before an actual mission is difficult if not impossible. Simulated images are needed to prototype, implement and validate Image Processing (IP) algorithms in preparation for space exploration missions. Several simulation engines have been developed and used worldwide, such as PANGU [2], VEROSIM, Blender, OSGEarth, Unreal Engine or SISPO [3]. At Airbus we developed the SurRender software with a list of desirable features in mind. The objective was to cover mission development cycles from preliminary analysis and sizing to advanced design, development and validation phases. Key requirements include:

- Interfaces with data formats such as NASA PDS: Digital Elevation Models (DEM): .img, albedo maps with many standard image formats: .png, .tiff, .jpeg2000, etc. and 3D meshes: .obj (remark: computer graphics formats are needed, not CAD models.)

- Raytracing shall be available for high fidelity simulations.

- Real-time rendering shall be achievable for integration in closed-loop simulation environments.

- Computation must be performed in double precision (float64) and image sampling must be optimized to manage needed dynamic range (from millions of km to contact).

- The simulator shall offer high flexibility to modify models for sensors, materials, etc.

- The simulator shall optimize memory management to allow large datasets.

- Images shall be validated geometrically and radiometrically.

## 2.2 Why is SurRender performance unique?

SurRender uses standard computer graphics concepts (scene graphs, bounding boxes, shaders, etc.) but it is based on a proprietary implementation of many functions. There are two concurrent pipelines: a (tuned) OpenGL pipeline for real time applications and an original raytracing pipeline for IP development and performance assessment. Most rendering engines (Unreal Engine, OSGEarth, etc.) use OpenGL, a 3D graphics standard widely used in the video game or animation industry that benefits from hardware-accelerated rendering (GPU). It has some drawbacks. For instance reference [4] showed the limits of the rasterization techniques to simulate an instrument PSF (Point Spread Function). Noises and sensor models can only be implemented as a post-processing and have limited representativeness. OpenGL implements the principle of far-plane / near-plane (background / foreground) which is unphysical and may yield numerical precision problems. In contrast SurRender splits the scene in an optimal number of layers even in OpenGL. As we will see double precision is essential. It is only locally implemented in OpenGL (eg. for quaternions) thus OpenGL simulations are not always numerically reliable. Intrinsically it does not allow going beyond simple projection models (pinhole): this is a limitation of rasterization which requires the projection of triangles to be triangles.

With the increase of computing performance, general purpose rendering engines are starting to implement raytracing techniques on GPU. However using raytracing does not necessarily mean having a physical representativeness. Tricks are used to make the image look visually appealing, for instance images may be subsampled for rendering before being oversampled with a neural network. Blender offers interesting raytracing capabilities; in particular the Cycles engine is able to simulate simple camera effects. SISPO is based on Blender Cycles, it offers specialized features targeting scientific space applications and has demonstrated good performances. Blender EVE engine literally "renders what the eye can see"; it is designed to trick the human eye. A strong limitation is RAM management. For Blender for instance, elevation models are converted to 3D meshes which quickly saturate RAM. In contrast SurRender uses an in-house format for DEM which are stored as conemaps (representing local tangents) and heightmaps (relative elevation). The data are initialized in mass memory using memory mapping and only the needed details are loaded in RAM. Furthermore this alleviates limits on the size of datasets: for instance the entire Moon can be covered in a single dataset without precision loss. In general only part of a meshed object is loaded in RAM (in video games this is the classical situation where objects appear successively with some lag), and this is not compatible with real-time for system validation. SurRender uses its internal representation for the graph scene that guarantees correct rendering even in OpenGL.

A (backward) raytracer does not only sample the detector, it samples 3D space. The scene is sampled with rays and a large enough number of rays must be cast from the pixel plane to obtain sufficient (numerical) signal-to-noise ratio (SNR). The sparsity of scenes and large scale ranges call for specialized methods. Without special optimization a raytracer would be highly inefficient as it would dedicate most of the resources to sample empty space. In raytracing if an object is not targeted explicitly it is unlikely it will be sampled. In a worst case scenario, the renderer may never intersect a distant object because it represents a very small solid angle. SurRender implements its own raytracer on CPU in full double precision. It is designed to sample space where it matters. First the rays target the bounding sphere of objects (preferential sampling). Second, the rays are used efficiently: the renderer targets in priority subparts of the scene responsible for the highest variance (importance sampling). There is more weight in regions with more signal. In particular, rather than uniform sampling, the raytracer uses the density function of the PSF with optimal statistical estimators. This way the physics of the light rays (optics diffraction) is intrinsically simulated rather than relying on a post-processing. A dichotomy is performed on the PSF such that these principles are applied at the subpixel level. They guarantee that a very high image quality is achieved with less rays. SurRender intrinsic use of PSF models for ray sampling (as opposed to post-processing) guarantees radiometric and geometric accuracy at the subpixel level.

Thanks to these optimizations, we can render complex scenes in raytracing in a sizeable timeframe. Examples include for instance secondary illumination from a planet on a spacecraft, continuous simulation of far away objects with constant level of noise from "infinity" to contact: a distant object may be unresolved but it radiometric budget is still a function of its apparent size. SurRender also has specialized routines for the rendering of stars which is systematically overlooked by other engines. Something that is very important for instance for Lunar missions, is that SurRender renders shadows physically, even in OpenGL for which it does not rely on the standard shadow mapping technique when rendering elevation models (or other kind of analytical models). Soft shadows account for direct illumination from the Sun - which is modelled as an extended source -, secondary illumination from other bodies (Earth, etc.) and from the local terrain, and self-reflections from the spacecraft itself.

Numerical precision is a key requirement for space applications because the dynamic range between celestial distances ($>>10^7$ m) and details of the target objects (satellite, local height, ~mm precision) is not compatible with the dynamic range of 32 bit floats (~$10^9$). Furthermore in some cases it is necessary to render each individual pixel along a pixel row sequentially in which case the simulation needs to be done in the time domain. For instance, regarding rolling shutter, push-broom or LiDAR sensors implementations, one must take into account the photons optical path and the camera relative motion during target acquisition. Other optimizations exist at lower level. For instance SurRender has its own implementation of all mathematical functions, which warrants a good tradeoff between speed and precision, and control over the compiler between different platforms.

Finally it is essential for a simulator to be flexible. It means to have a modular architecture that is compatible with numerous new models inputs. In SurRender all models can be tuned using SuMoL (SurRender Modelling Language) and dedicated interface. Some examples of original models particularly useful for space applications include relevant material surface properties (BRDF from Hapke, Oren Nayar and more), analytical shapes (spheroids, etc.), pointing error models, variable PSF models, various projection models, etc. SuMoL models are compiled "on the fly" by the engine. This makes it possible to do sensitivity analysis by varying all parameters from the simulation. This is a precious asset to do IP performance assessment for advanced mission phases.

## 2.3 SurRender 7: a major release for the software 10th anniversary

For its 10th anniversary, the SurRender team has upgraded the software to an industrial level. The software has undergone a complete refactoring and it now offers faster OpenGL and raytracing engines. Improvements include optimizations for new use cases such as rover applications, new functions for loading and positioning lists of 3D meshes, LiDAR support, direct rendering to HDMI ports, a lightweight C interface, better statistics estimators and cloud computing features. An industrialized CICD process has been implemented based on Airbus best practices for satellite ground segments and more tests have complemented SurRender validation reports. Following up on feedbacks from SurRender growing user list, the user experience has been improved. The API and the server log became more user-friendly and clear. Additional tools such as those used to preprocess PDS datasets have been upgraded. A graphical demonstration interface will be released with SurRender 7 to simplify the discovery of the tool by new users. SurRender software is a professional software and licences can be granted on the basis of commercial or academical licences.

## 3 RENDERING OF PLANETARY SURFACES

### 3.1 The entire Moon in a single dataset

Brochard et al. [1] first presented a simulation of the Moon used for the development and validation of IP algorithms for planetary descent and landing. It is based on publicly available data: a global Digital Elevation Model from Lunar Reconnaissance Orbiter / Lunar Orbiter Laser Altimeter (LRO / LOLA) at 118m resolution (GSD: Ground Sampled Distance), and an albedo map from JAXA SELENE / Kaguya Multiband Imager at 237 m [5,6]. Dataset of better resolution exist, although not at a global scale. For instance LRO / LOLA DEM are available with a GSD of 59 m but only at latitude ranging from -60° and +60° so the dataset is not adequate for instance for South Pole missions. LRO offers imaging down to 1 m resolution locally. Our simulator has the benefit of covering the entire Moon with a single set of data (38 GB in compact format). RAM is managed very efficiently: the resolution is adapted to perform continuous simulations from tens of thousands of km to touchdown, anywhere on the surface. SurRender can use datasets up to 256 TB. Raytracing and pathtracing render realistic shadows accounting for occlusions, secondary illumination and Sun solid angle. The terrain optical properties are as good as reference models allow. For example the community standard for regolith surfaces is the Hapke BRDF (Bidirectional Reflectance Distribution Function) which captures the zero-phase situation when a pixel LoS is aligned with the Sun direction (opposition surge), drastically reducing contrast. Initial simulations were made with a sampling of 128 rays/pixel for high quality rendering. We can trade-off image quality for computing performance: with optimized numerical parameters, the simulation runs at 5 Hz in raytracing on a modern workstation with quality level adequate for real time campaigns (SurRender efficient raytracer yield good SNR with few rays/pixel).

In Figure 1, we show a validation test that was carried out independently from our team to verify the geospatial correctness of the Moon images. Our dataset was reproduced and images were compared with the *Lunar Crater Database* which provides a census of 1.3 millions craters [7] (the accuracy and completeness of this reference database is on a best-effort basis). Visual comparison shows that the image is correctly georeferenced and projected – here with a pinhole model and 70° field-of-view (FOV). To support future Lunar missions, the simulator will soon be upgraded with higher resolution datasets.

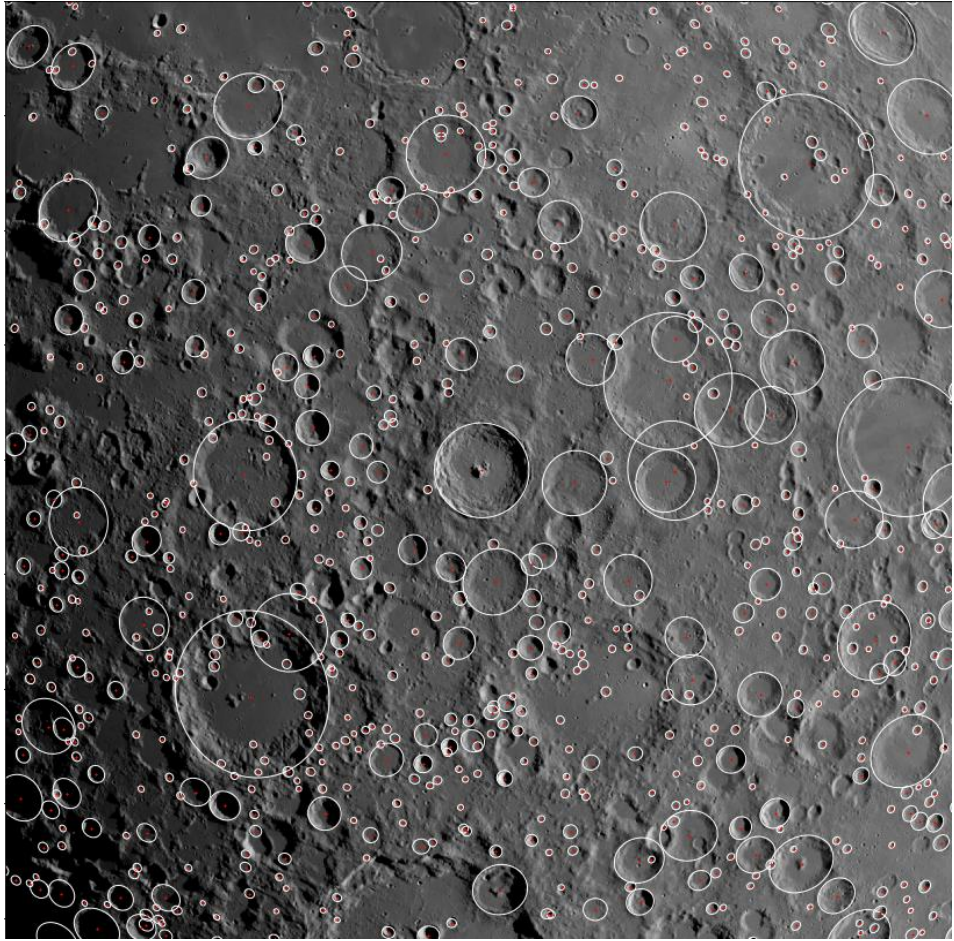

Figure 1. Craters from the *Lunar Crater Database* superimposed on a SurRender rendering of the Moon centered on the Tycho crater at 1000 km altitude.
(courtesy of Wouter Doppenberg, TU Delft)

### 3.2 Asteroids: meshes vs spherical DEM

SurRender can be used to simulate small solar system bodies to develop reliable Image Processing (IP) solutions such as 3D reconstruction, hazard detection and navigation. For instance reference [8] presents relative navigation and 3D reconstruction techniques tested on SurRender images. The achievable rendering quality depends on the input dataset representativeness. The philosophy of SurRender is to take as input real datasets rather than generating synthetic 3D models. Asteroids can be represented classically by 3D meshes, or by SurRender spherical (or planar) DEM format providing the geometry is close enough to a spheroid. In Figure 2 we show three small bodies rendered either with 3D meshes (.obj) or a spherical DEM. The first two examples are Itokawa and Vesta which were modelled using 3D meshes for real poses from Hayabusa and Dawn mission data [9, 10].

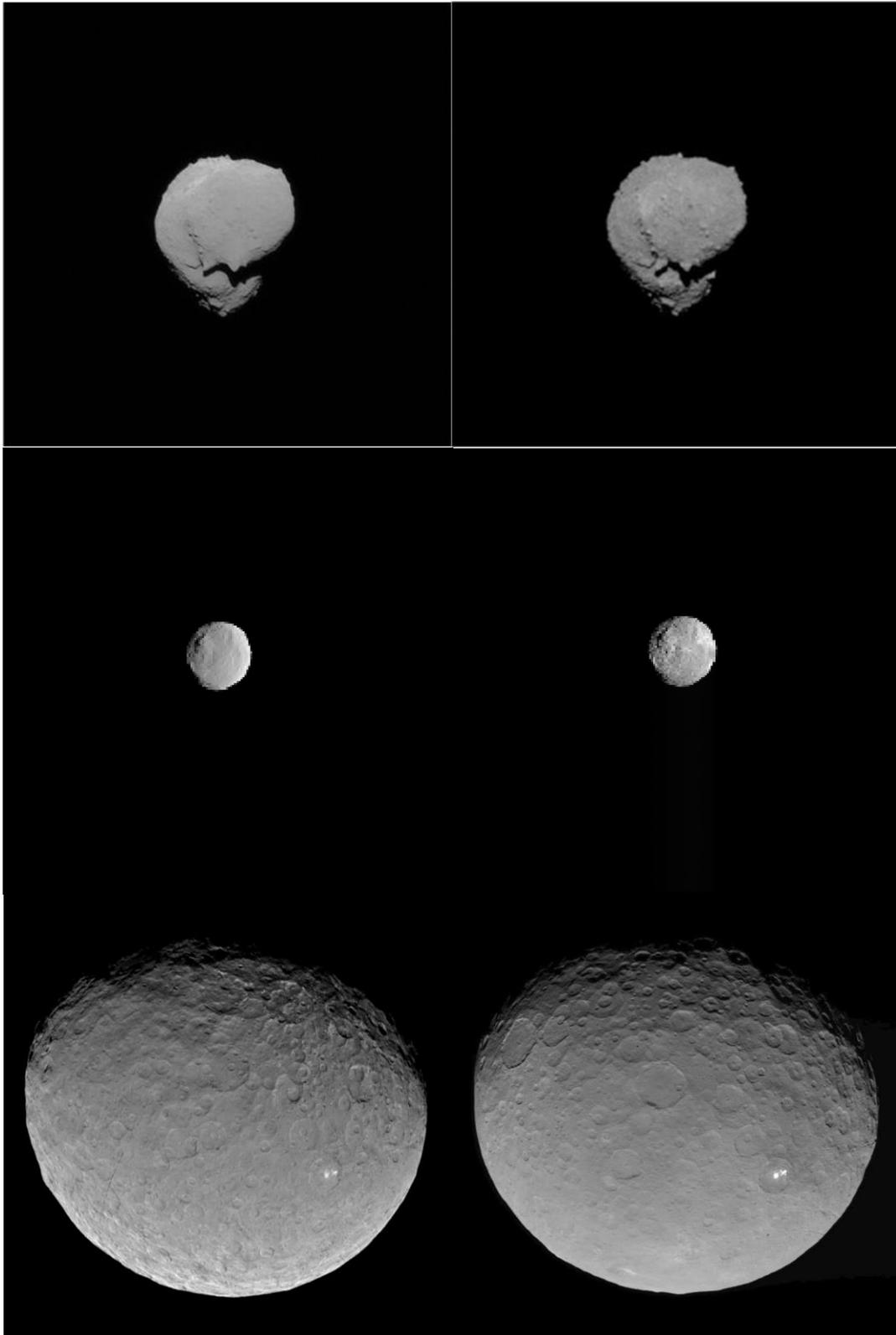

Figure 2. Small Solar System bodies rendered with SurRender (left) vs real images (right).
Top: Itokawa simulated with a 3D mesh (obj file) and a model of the Hayabusa/AMICA camera.
Middle: Vesta seen by Dawn RC3 camera and modelled with a 3D mesh; in both cases ephemeris
are taken from mission data (SPICE). Bottom: Ceres as seen by Dawn RC3 Camera and simulated
with a spherical DEM model; in this case the scene geometry is approximate.

*ESA GNC 2021 – J. Lebreton* 7

Visual comparison with real images from the AMICA and RC3 cameras respectively is compelling, the only noticeable difference is that the albedo maps and DEM resolution slightly reduce the granularity of the terrain. Let us remark that the position of the target is shifted with respect to the reference image. This is a common problem due to the fact that the metadata associated with the image are not infinitely accurate: poses are given by IMU and radar positioning and they are accurate only to a fraction of degree which amounts to several pixels in this 5 degree FoV image.

In the last row of Figure 2, a view of dwarf planet Ceres is shown. This rendering is based on complete mosaic and DEM from Dawn Framing Cameras at ~140 m resolution and 10 m vertical accuracy. Ceres radius is 470 km and the 3D model corresponds to >200 million vertices. A DEM is much more compact than a mesh (25x) and supports varying levels of details. Only details which have an impact on the final image are accessed. They do not suffer from precision issues for large objects (meshes are usually stored in single precision only). Furthermore there are no memory access delays after initialization of the mass memory (memory mapping). A DEM can be bigger than the actual amount of RAM on the computer with little impact on rendering speed. The high quality of the dataset yields impressive rendering quality very representative of the real image. In all three cases it is important to note that the rendering were made directly on the PDS data in the absence of any correction of input or post-processing.

### 3.3 Physical LiDAR model for Mars landing

SurRender includes a physics-based simulator of LiDAR cameras (flash LiDAR). Rather than relying simply on depth maps as it is the case in traditional approaches, the raytracer accounts for the light propagation time. A light beam is emitted and rebounds on the environment. The return signal is integrated in each pixel using either phase shift detection or time delay measurements to calculate the range. The advantage of the approach is that it benefits from SurRender simulation framework so it can accounts for the detector physics, for motion blur, for surface reflectance models, etc. In Figure 3 we illustrate the performance on a Martian terrain using a 1 m resolution DEM and albedo map from MRO / HiRISE [11] (credit: NASA / JPL / U. Arizona). The LiDAR implemented here is a toy-model of a time-of-flight camera based on phase shift measurement.

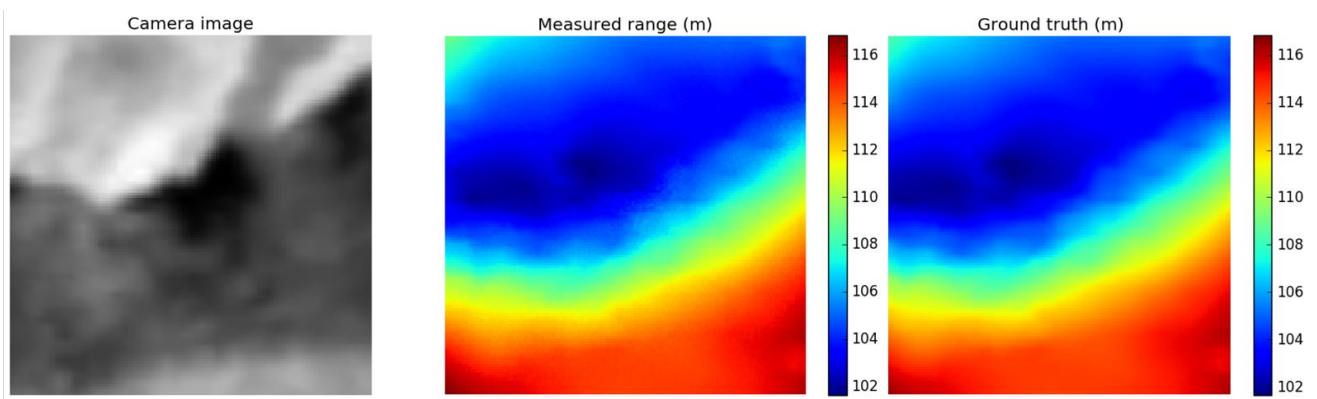

Figure 3. LiDAR camera view of a Martian landscape (West Candor Chasma, left). The range is estimated with a physics-simulation of the wave-front (middle). The depth map serves as the reference ground truth (right).

### 3.4 Synthetic view of the Jezero crater

Since SurRender 7, we optimized the rendering engine and added new functions to satisfy the requirements of rover applications. Key differences with space-based imaging are for instance typical distances and the presence of the sky background. The sky is represented by a texture on a sphere surrounding the planet (atmosphere models are not available to date). The sky is an important contributor to the illumination of the scene in addition to primary light source(s) and secondary illumination. It is sampled by the raytracer like other objects. Figure 4 shows a simulation of Martian terrain. The DEM is the Jezero crater as measured by MRO [11]. Additional details were added using a Perlin generator. The albedo map is an ad-hoc sandy terrain texture and a Hapke BRDF is used. A simple model of the Exomars rover is placed in the center (courtesy of ADS UK). Thousands of (identical) 3D mesh models of rocks are projected on the surface with random orientations and scales. The engine was tested in two concurrent setups: OpenGL simulation for real-time tests, and high quality raytracing for detailed IP design and validation.

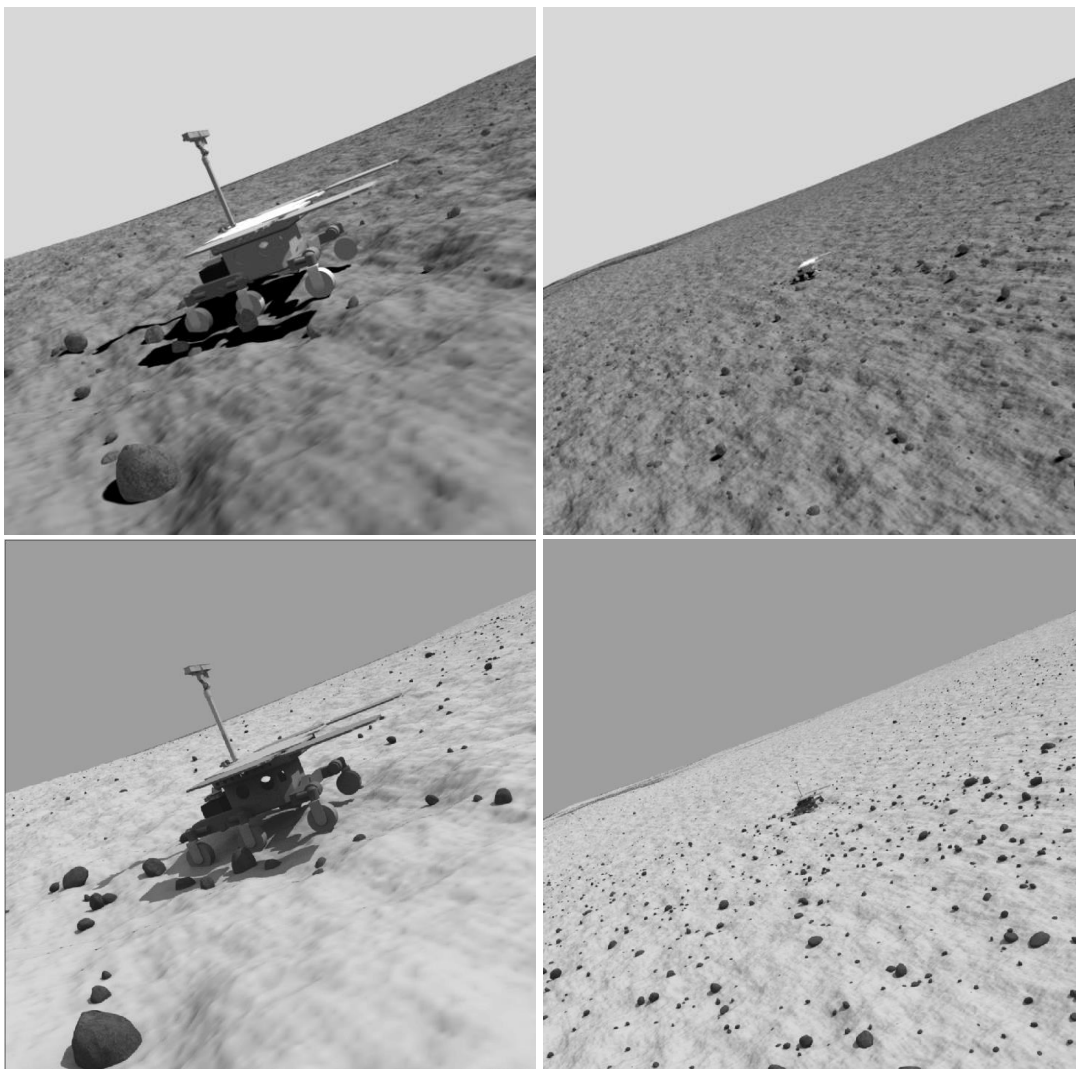

Figure 4. A simple model of Exomars rover in Mars Jezero crater. Synthetic details were added to the DEM and thousands of rocks are randomly distributed on the terrain. The sky is rendered with a uniform texture (no clouds). Top panel: OpenGL. Bottom panel: raytracing. (colorbars are different)

In both cases, one point of attention is the quality of soft shadows rendered by the multiple illumination sources. In OpenGL the dark side of rocks and terrain below the rover are completely black, whereas in RT they are illuminated by the environment. The sharpness profile of object edges is obvious in raytracing. These details would impact IP filters results. 1024x1024 pixels images were rendered, from a 300m x 300m DEM at 4cm resolution, a list of 12,000 3D meshes and a PSF. The real-time (OpenGL) simulations run as fast as 9Hz on a laptop (Quadro T1000 GPU, 2.5GHz Core i5-9400H processor). In contrast, the high quality images (64 rays/pixel) were rendered in about 300 seconds. The sky contributes greatly to the global illumination but this effect cannot be modeled in OpenGL. The color scale is not the same between the two simulations.

## 4 RENDEZVOUS WITH ARTIFICIAL OBJECTS

### 4.1 Earth satellites and MSR ERO Orbiting Sample

SurRender was initially developed for in-orbit rendezvous applications. As always the quality of any simulation depends on the complexity of input models. In Figure 5, top left panel, a rather sophisticated 3D model of satellite SPOT 5 is simulated on an Earth background. In the top right panel, a simulation used for the development of RemoveDEBRIS [12] VBN algorithms is shown.

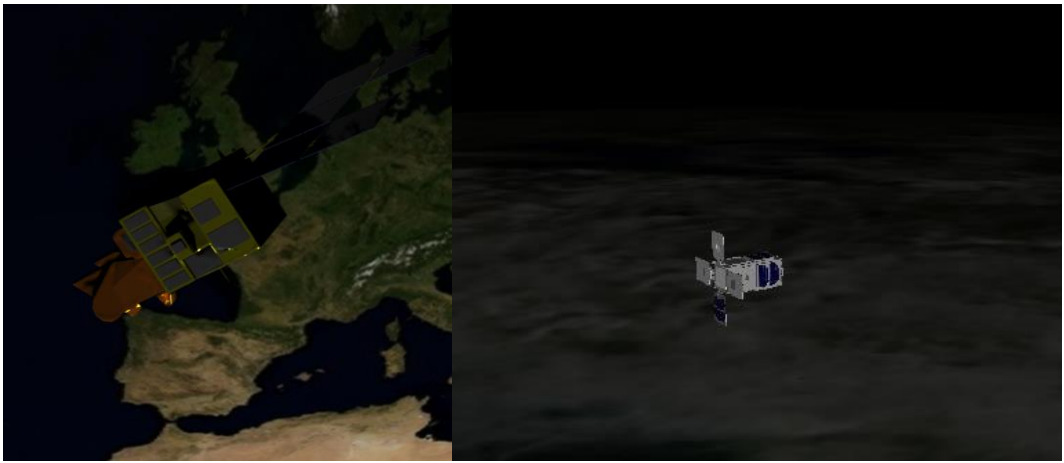

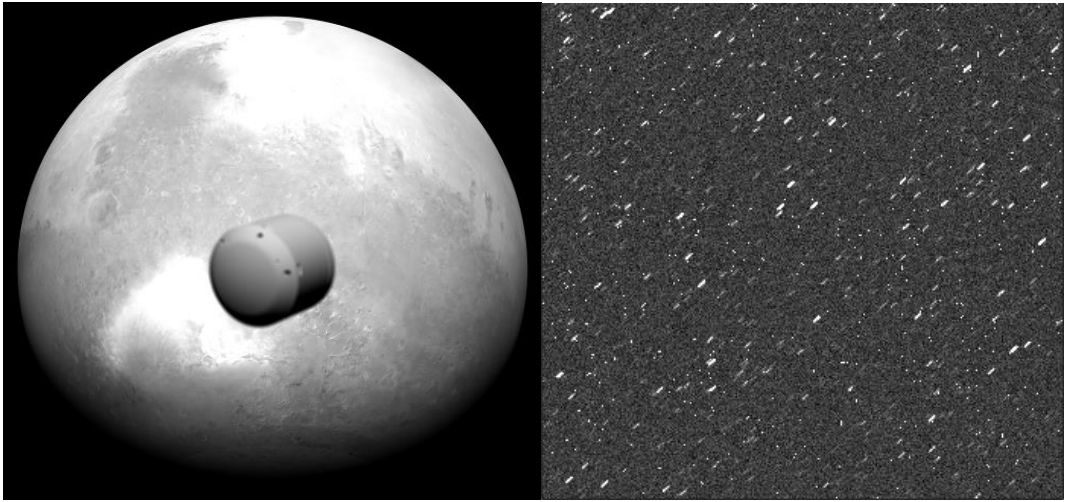

Figure 5. Top: The SPOT 5 satellite and the RemoveDebris demonstrator on an Earth background.
Bottom: Examples of preliminary SurRender simulations performed for MSR ERO.
Left: at close range. Right: at far range on a background of stars and radiations.
The scales are exaggerated for illustration.

Our team develops the detection and rendezvous algorithms for the MSR ERO mission. In Figure 5, bottom panel we show preliminary simulations rendered with SurRender. The models include an advanced sensor and secondary illumination from Mars. Information about the synthetic images and their counterpart obtained on a robotic test bench can be found in [13]. Stars and radiations are critical confusion sources for distant target detection, and it is essential to simulate realistically their aspect including in particular the effect of vibrations, motion blur and defocus. The radiation simulator is an external module. Standard star catalogues (e.g. Tycho-2) can be read by SurRender in csv or binary format and rendered with proven geometrical and radiometrical precision.

### 4.2 Revisiting LIRIS datasets for validation: approach of the ATV to the ISS

Real space images with reliable ground truth are rare. The LIRIS project - an experiment that flew on ATV-5 to test VBN solutions – represents an important heritage. The trajectory was decomposed in a fly-under phase with distances ranging from 70 to 8.8 km, and a rendezvous phase from 30 km to contact. LIRIS embarked a visible camera, two infrared cameras and a LiDAR. We focus here on the visible camera which can be modelled with SurRender using available datasheets. It has a 57.5 x 44.9° FoV, and a resolution of 1360 x 1024 pixels and a focal length of 8 mm. The images were distortion-corrected using polynomial models from the camera supplier. After this correction the resolution is 1437 x 1079 pixels. Ground-truth poses are available for the full sequence: the relative state vector was estimated using available navigation sensors depending on mission phase (RGPS, radio navigation, gyrometer, star-tracker). It is essential to point out that the ground-truth has limited accuracy: LIRIS user manual (internal) reports errors and biases not better than 6m to 30m in the rendezvous and fly-under phases respectively. The rendezvous phase was analyzed by [14] who tested model-based tracking performances. We revisit their models and processed datasets.

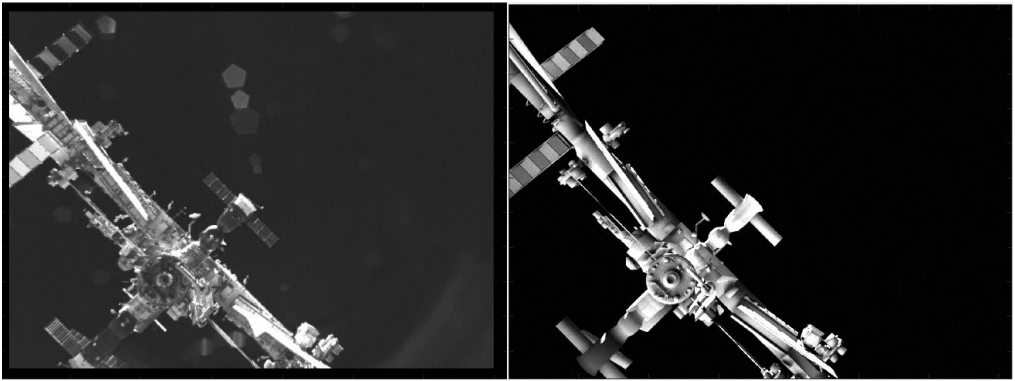

Figure 6. A LIRIS image (left) and the corresponding simulated image (right) during the rendezvous phase. The focus of this simulation is geometry and relative radiometry, there is neither a detector model (in particular no lens flare, no fixed pattern noise) nor textures on the ISS here. However projection models and secondary illumination from Earth are well represented.

Available sensor models are complemented by evaluations made directly on the image. The PSF is assumed Gaussian and its width is measured on distant images when the size of ISS is smaller than one pixel. Due to the lack of information on the optical chain (collective surface, transmittance, quantum efficiency) and integration time, we have to evaluate the gain (e-/irradiance) and noise budget. The gain is obtained by comparing the signal measured in LIRIS images and the signal in simulated irradiance images on a series of images. Additive noise is evaluated from the background and is composed of a constant term and a white noise term. We perform simulations matching LIRIS scenarios with the specified geometry for relevant objects (Sun, ISS, camera, Earth) and a 3D model of the ISS. The Sun is the primary light source on the ISS model, but secondary light from Earth proves an important contributor. The sun power is integrated on a spectral band 300-1100 nm in the sensor.

Figure 6 shows the nominal setup. There is no sensor model in this simulation which explains the black background; the LIRIS camera has a fixed pattern noise and strong lens flare in this case. At close-range, the ISS occupies a large fraction of the detector. The global illumination of the ISS is very consistent between the simulation and the real image, it includes an important contribution from Earthshine. The global geometry of the ISS projected in the image plane is correct as shown by the good performances of VBN algorithms [14]. Of course the 3D model is simplified, and it does not include representative textures (albedo maps) for elements such as solar panels.

At large distances, details of the texture do not matter for the global radiometric budget. We focus now on the far-range phases of the mission when the target is barely resolved. Figure 7 shows that it is impossible to differentiate simulations from real images in this case. We illustrate the simulation results at 3 distances: 59, 31 and 22 km. At these distances, the ISS represents about 3 to 9 pixels on the detector plane. The figures show that for such a distant object, the geometry and the radiometry simulated by SurRender are perfectly representative of the real images. This analysis demonstrates that SurRender simulations are only limited by the representativeness of the input models. Provided that the input data are understood and used wisely, SurRender produces reliable simulations for IP development and validation.

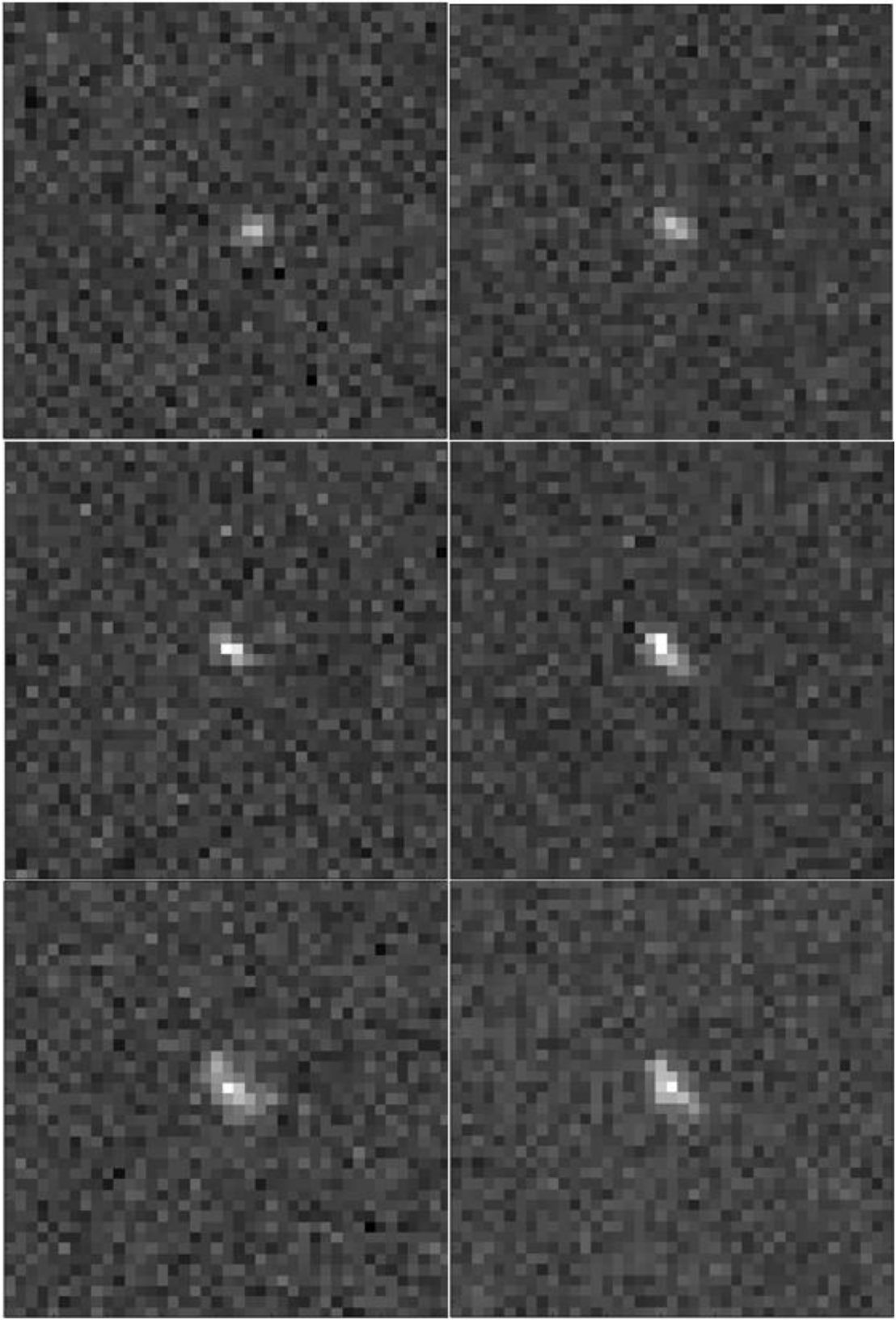

Figure 7. Real images (left) and simulated images (right) of the ISS at 59 km, 31 km and 22 km from the camera (from top to bottom). Regions of interest of size 41x41 pixels and centered on the position of ISS are displayed. The gray scale bar is the same for all images (100-300 LSB).

# 5 END-TO-END SIMULATOR FOR VALIDATION CAMPAIGNS

SurRender software is fully integrated in the image chain development and validation process for several missions built by Airbus for ESA. To date the most mature example is JUICE. The limb-matching navigation algorithm (EAGLE IP) relies on representative simulations of Ganymede, Callisto and Europa [15]. SurRender simulates in great amount of details the spacecraft navigation camera. It is used at IP level for detailed performance assessment based on numerical models, for open-loop simulation and for closed-loop simulations on the spacecraft engineering model with the actual camera. Each use case has its own requirements and various interfaces and configurations are used (Python, Matlab, Simulink, real-time C interface). The interested reader can refer to [16, 17] for details about EAGLE IP. Figure 8 presents a simplified camera model not representative of the actual mission model. This figure illustrates that for IP performance campaigns, SurRender can be used to systematically disperse all uncertain parameters. In this example, three high-level parameters are dispersed. This is made possible thanks to SurRender flexible architecture, to SuMoL models and distributed computing capabilities.

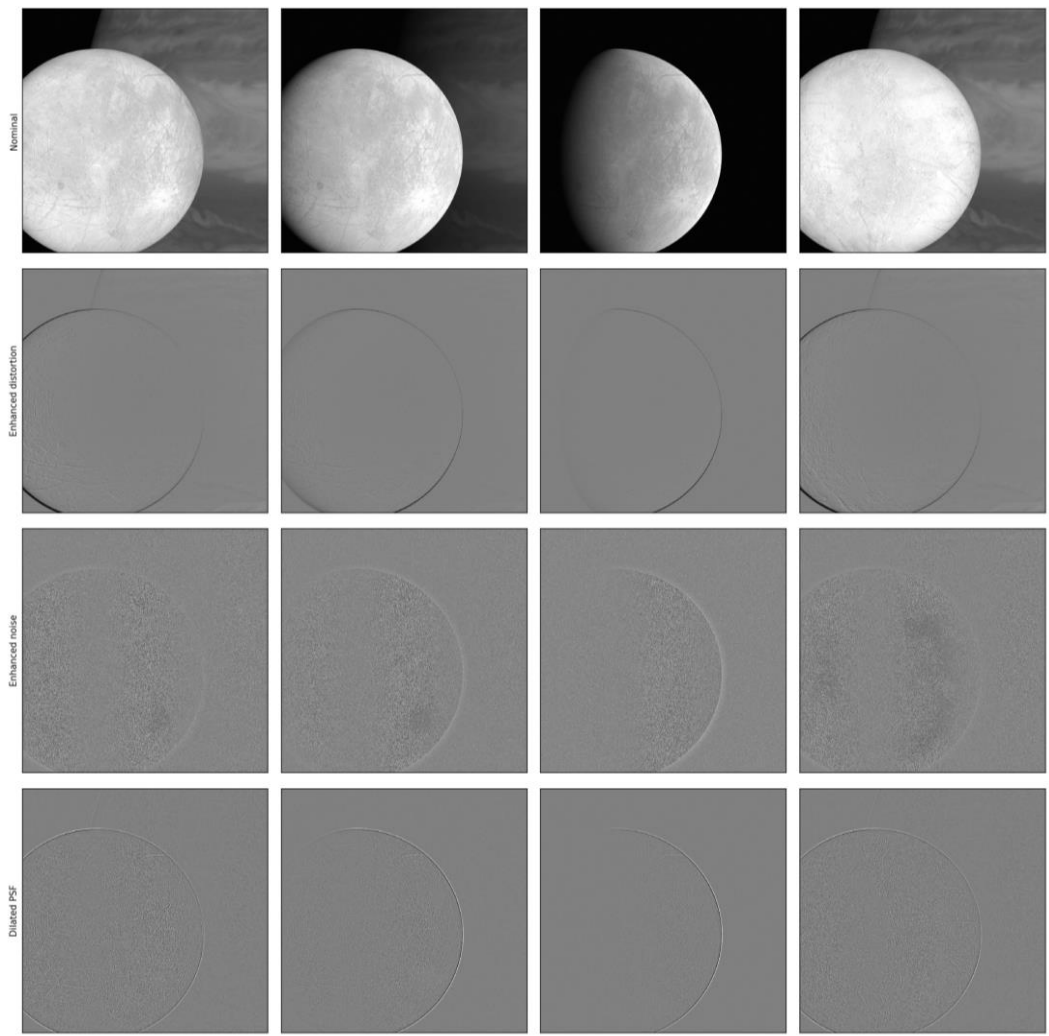

Figure 8. Examples of rendering of Europa performed with a simplified JUICE simulator (not representative of the actual NavCam model). From left to right different geometries are explored. On the top row is the camera nominal model. On the next rows the difference with nominal images is shown when some model parameters are dispersed: respectively the distortion model, the noise model and the PSF model.

# 6  CONCLUSIONS

The SurRender software is a very powerful image simulator that covers the particular requirements of space mission IP design, development and validation. This paper highlights some of its achievements on use cases such as exploration of the Moon and Solar System objects and space robotics. Some validation proofs are provided that complement existing validation reports based on formal demonstration and comparisons with real data. SurRender is already used by multiple users outside of Airbus in academia and industries worldwide. A major release of SurRender 7 will be advertised shortly after the ESA GNC conference. As a next step, our team is planning to upgrade existing tools used for input datasets enhancement, such as fractal detail generation for textures and DEM, and procedural loading of 3D models (boulders). New use cases are also emerging in the field of earth observation and rovers and dedicated features are under development.

# 7  COPYRIGHT